\documentclass[onecolumn,a4paper,12pt,amsmath,preprintnumbers,nofootinbib,tightenlines,aps]{revtex4}\pdfoutput=1
\usepackage[english]{babel}
\usepackage{bbm}
\usepackage{graphicx}
\usepackage{hyperref}

\graphicspath{{pics/}}

\newcommand{\hh}[1]{\left(#1\right)}
\newcommand{\bb}[1]{\big[#1\big]}
\newcommand{\bhh}[1]{\big(#1\big)}

\newcommand{\norm}[1]{\lVert#1\rVert}

\newcommand{\RR}{\mathbbm{R}}

\newcommand{\ii}{\mathbbm{i}}

\newcommand{\md}{\mathrm{d}}

\newcommand{\cs}{\mathcal{M}}
\newcommand{\csp}{\bar{\mathcal{M}}}
\newcommand{\hf}{\tfrac{1}{2}}

\newcommand{\s}[1]{\mspace{#1mu}}

\newcommand{\lp}{\textrm{lam}}

\hypersetup{pdfauthor={Maarten van de Meent},pdftitle={Gravitational Waves from Piecewise Flat Gravity},pdfkeywords={ Locally flat - Straight strings - Classical general relativity - Weak field approximation - Gravitational Waves}}

\begin{document}
\preprint{ITP-UU-10/46}
\preprint{SPIN-10/39}
\title{Piecewise Flat Gravitational Waves}

\author{Maarten \surname{van de Meent}}
\affiliation{Institute for Theoretical Physics and Spinoza Institute,\\ Utrecht University,\\ P.O. Box 80.195, 3508 TD Utrecht, the Netherlands}
\email{M.vandeMeent@uu.nl}

\date{\today}

\begin{abstract}
We examine the continuum limit of the piecewise flat locally finite gravity model introduced by 't Hooft. In the linear weak field limit we find the energy--momentum tensor and metric perturbation of an arbitrary configuration of defects. The energy--momentum turns out to be restricted to satisfy certain conditions. The metric perturbation is mostly fixed by the energy--momentum except for its lightlike modes which reproduce linear gravitational waves, despite no such waves being present at the microscopic level.
\end{abstract}
\maketitle

\section{Introduction}
Einsteinian gravity in $2+1$ dimensions is in many ways much simpler than in $3+1$ dimensions. In $2+1$ dimensions the Einstein equation completely fixes the Riemann tensor in terms of the energy--momentum. In particular, there are no local gravitational degrees of freedom.\cite{Deser:1983tn} This makes quantization of gravity in $2+1$ dimensions possible as was done non-perturbatively by Witten in 1988 \cite{Witten:1988hc} (perturbative quantization was shown the following year by Deser, McCarthy, and Yang \cite{Deser:1989}). Local degrees of freedom can only be added by including matter. A system of point particles interacting gravitationally has a locally finite number of degrees of freedom and may be quantized.\cite{'tHooft:1993nj,Waelbroeck:1994iy, 'tHooft:1996uc}

The geometry of a system of point particles in $2+1$ dimensions  is that of a piecewise flat manifold consisting of blocks of flat spacetime glued together along their faces.\cite{'tHooft:1992} All curvature is concentrated along the edges of the blocks, which coincide with the paths of the particles. In 2008, 't Hooft suggested \cite{hooft2008} that to obtain a locally finite theory of gravity in $3+1$ dimensions, one should impose the rule that empty space has no local structure and is therefore locally (Riemann) flat. The combination of this rule with Einstein's equation implies that the only local degrees of freedom that may be added are curvature defects of co-dimension 2, i.e. lines propagating at constant velocity. The result is a model that can be regarded as a generalization of gravity in $2+1$ dimensions, which we studied in more detail in \cite{meent:2010}, where it was shown that a configuration of defect lines in $3+1$ dimensions can be described as a piecewise flat manifold.

In this respect the model is very similar to the piecewise flat approach to gravity introduced by Regge in 1961.\cite{regge:1961} Regge calculus has since then been used as a tool both in classical numerical general relativity as well as in quantum gravity. (See \cite{williams:1992} and \cite{RW:2000} for a review and references.)  A key difference is that Regge calculus allows defects of any signature, whereas 't Hooft's model insists on interpreting the defects as propagating physical degrees of freedom. Imposing causality then implies that only non-spacelike defects may appear. 

In 't Hooft's model all physical degrees of freedom, including the gravitational ones, are defects and therefore treated as matter.  As a result, by design, the model has no local gravitational structure. This implies that --- a priori --- there are no long range gravitational fields and gravitational waves, unlike the world that we observe, and therefore raises the question how physical this model can be. A resolution may come from the fact that the model allows both positive and negative energy defects, i.e. defects with a positive or negative deficit angle respectively.\footnote{The appearance of negative energy degrees of freedom may appear unnatural. However, since in our treatment gravitational excitations are included in the energy--momentum side of Einstein's equation, they may appear with an opposite sign. Classical positivity conditions will, at best, only be valid on larger scales in this model.} One could imagine a configuration of defects that on large scales has zero average energy--momentum, but  has a non-trivial average Weyl curvature. In other words, we might hope that the vacuum structure found in general relativity is recovered in the continuum limit.

The continuum limit of piecewise flat gravity models has been studied extensively in the context of Regge calculus.\cite{CMS:1982,CMS:1984,Feinberg1984343,Friedberg1984145} It is known that the space of Regge manifolds is dense in the space of solutions of general relativity.  In particular, Regge calculus contains approximations to any vacuum solution of general relativity. However, this does not answer our question since we restrict to piecewise flat configurations that only contain physical (i.e. non-spacelike) defects. It is not clear whether this subset of configurations is still dense. This question will be addressed in this article.

Our primary interest in this model has been to handle it as a precursor to a possible theory of quantum gravity. Just as $2+1$ dimensional gravity, it may teach us interesting lessons about the quantization of gravity, regardless whether it serves well as a model for real world physics. The emphasis on causality and the fact that the studied configurations are exact solutions of general relativity, may however also make this model interesting as an approximation scheme for classical inquiries in general relativity. For example, it would be interesting to see if it could be used as a test case for questions about the averaging problem in cosmology. For such applications it is crucial to know what kind of matter sources may be approximated. This will be a second line of inquiry.

The description of a configuration of defects simplifies dramatically in the limit where the energies of all defects are considered to be infinitesimal. This allows us to study continuous distributions of infinitesimal defects. Section \ref{sec:configs} explains how to describe such a configuration in this limit using a density function on the configuration space of an individual defect and introduces the notation used in the rest of the article.

Section \ref{sec:T} then constructs the energy--momentum tensor produced by an arbitrary configuration of physical defects. It finds the conditions that the energy--momentum will obey (and consequently must be obeyed by any theory that is to be approximated by this model in the limit of weak fields).

The metric perturbation produced by a general configuration of physical defects is obtained in section \ref{sec:h}. After which, in section \ref{sec:gravwave}, the results of the preceding sections are combined to find the metric perturbations that can be produced by a configuration with vanishing energy--momentum. We find that it is possible to reproduce the complete spectrum of gravitational waves found in linearized Einstein gravity.

\section{Defect configurations in weak field}\label{sec:configs}
In the previous articles \cite{hooft2008,meent:2010} two approaches to describing a general configuration of propagating defects were described.  Both of which share that they rapidly increase in complexity as the number of defects increases, because the description of the state of a defect involves the states of other defects as well. This convolution of the description is this model's manifestation of the non-linear nature of general relativity. 
 
However, in this paper we want to look at geometries generated by configurations with large numbers of defects, which makes the previously employed methods prohibitively complex. Fortunately, the description of a configuration of defects drastically simplifies in the limit that all defect angles are very small. In this limit each defect can be treated as a linear perturbation to a Minkowski background. Moreover, any new intermediate defects that would be created by the collision of two defects are higher order in the defect angles of the colliding defects and can be neglected.

Consequently, in the linear limit the state of a defect line can be described while ignoring the presence of other defects. A configuration of defects can therefore be completely described by giving the number of defects in any given state. Since the effect of two defects with the same state is simply that of a single defect with the combined energy of the two defects, we can completely describe the configuration by giving the energy in each possible state. That is, if $\cs$ is the state space of a single defect with unit energy density, a configuration of defects can be described by a distribution $\rho:\cs\rightarrow\RR$ giving the energy density in each state.

To parameterize the state space $\cs$, recall that in \cite{hooft2008} it was shown that the state of a single defect  line with a density $\rho$ could be described by the following data: a vector $\vec{p}$ that gives the position of the defect, a vector $\vec{d}$ that gives its direction, and a vector $\vec{v}$ that gives its velocity.

The triple  $\bhh{\vec{p},\hat{d},\vec{v}}$ is enough to uniquely identify the state of the defect. However, this characterization is not unique, since the triple  
\begin{equation}\label{eq:equivtriple}
\bhh{\vec{p}+\alpha\vec{d},\beta\vec{d},\vec{v}+\gamma\vec{d}}
\end{equation}
describes exactly the same state. In fact, the triples describing the same state as $\bhh{\vec{p},\vec{d},\vec{v}}$ can be completely parameterized by the numbers $\alpha$, $\beta$, and $\gamma$. The state space of a line defect with unit density $\cs$ is thus obtained from $\RR^3\times \RR^3\times\RR^3$ by modding out by the equivalence relation \eqref{eq:equivtriple}.
 
A set of unique representatives for each equivalence class in $\cs$ can be formed by taking a triple $\bhh{\vec{p},\hat{d},\vec{v}}$ that satisfies the following conditions,
\begin{equation}\label{eq:conds}
\begin{aligned}
\vec{v} &\perp \hat{d}, \\
\vec{p} &\perp \hat{d}, \text{and}\\
\norm{\hat{d}}&=1.
\end{aligned}
\end{equation}
Such a representative triple is unique up to a sign of $\hat{d}$, i.e. to get a unique representative $\hat{d}$ should be viewed as an element of the real projective plane, $\RR\mathrm{P}^2$.

In this article it will be convenient to decompose  density functions $\rho$ on $\cs$ in a set of canonical functions that we call \emph{laminar plane waves}. A laminar plane wave is a configuration where all defects have the same direction $\hat{d}_0$ and velocity $\vec{v}_0$ and the density $\rho$ is a plane wave function with wave vector $\vec{k}_0$ with respect to the position $\vec{p}$ of the defects. That is, the density function $\rho^{\lp}[\vec{k}_0,\hat{d}_0,\vec{v}_0]$ corresponding to a laminar plane wave  with wave vector $\vec{k}_0$, direction $\hat{d}_0$, and velocity $\vec{v}_0$ can be written as a function  of $\bhh{\vec{p},\hat{d},\vec{v}}$ representing an equivalence class in $\cs$ as follows,
\begin{equation}\label{eq:lpw}
\rho^{\lp}[\vec{k}_0,\hat{d}_0,\vec{v}_0](\vec{p},\hat{d},\vec{v})=
	e^{2\pi\ii \vec{k}_0\cdot\vec{p}}\delta(\hat{d}-\hat{d}_0)\delta(\vec{v}-\vec{v}_0)
.\footnote{The density function of course needs to be real. The expansion in complex exponentials is for the sake of convenience only. To obtain physical results we should consider only the real part.}
\end{equation}
Since $\vec{p}$ en $\vec{v}$ are always perpendicular to $\hat{d}$ it is sufficient to consider only laminar plane waves with $\vec{k}_0$ and $\vec{v}_0$ perpendicular to $\hat{d}_0$. The set $\csp$ of laminar plane waves with parameters $\bhh{\vec{k}_0,\hat{d}_0,\vec{v}_0}$ is complete in the sense that any density function $\rho(\vec{p},\hat{d},\vec{v})$ on $\cs$ can be written as
\begin{align}
\rho(\vec{p},\hat{d},\vec{v})
	&=	\int_{\csp} \md\vec{k}_0\md\hat{d}_0\md\vec{v}_0 \,
		\rho^{\lp}[\vec{k}_0,\hat{d}_0,\vec{v}_0](\vec{p},\hat{d},\vec{v})\bar\rho(\vec{k}_0,\hat{d}_0,\vec{v}_0) \\
	&=	\int_{\vec{k}\perp\hat{d}}\md\vec{k}\, e^{2\pi\ii \vec{k}\cdot\vec{p}}\bar\rho(\vec{k},\hat{d},\vec{v}),
\end{align}
where $\bar\rho(\vec{p},\hat{d},\vec{v})$ is a density function on $\csp$. The function $\bar\rho(\vec{k},\hat{d},\vec{v})$ can therefore be viewed as a partial Fourier transform of $\rho(\vec{p},\hat{d},\vec{v})$.\footnote{Conversely, $\bar\rho(\vec{k},\hat{d},\vec{v})$ may be obtained from $\rho(\vec{p},\hat{d},\vec{v})$ by an inverse partial Fourier transform.}

It will also be convenient to split the parameter $\vec{v}$ in a component collinear with $\vec{k}$ and a component perpendicular to both $\vec{k}$ and $\hat{d}$,
\begin{equation}\label{eq:vsplit}
\vec{v} = \omega\frac{ \vec{k}}{\vec{k}^2} + v\frac{ \vec{k}\times\hat{d}}{\vec{k}^2},
\end{equation}
where the ambiguous direction of $\hat{d}$ is chosen such that $v$ is non-negative. The usefulness of this split becomes apparent when we do a Lorentz transform. The Lorentz transform of a laminar plane wave with parameters $\bhh{\vec{k},\hat{d},\omega,v}$ is again a laminar plane wave, but with different parameters. It can be shown that the combination $k_\mu = (\omega, \vec{k})$ transforms as a 4-vector under the Lorentz transformation. In section \ref{sec:T} we will see that this is the wave vector of the corresponding energy--momentum tensor. In the remainder of this article will denote the parameters of a laminar plane wave as $\bhh{k_\mu,\hat{d},v}$. 

The physicality condition that we impose on the defects implies that the velocity of each defect must be smaller than or equal to $c=1$.\footnote{Throughout this article we use natural units such that $c = \hbar = 8\pi G =1$ and metric signature $(- + + +)$.} By squaring equation \eqref{eq:vsplit} we see that this implies that
\begin{equation}
\frac{\omega^2 + v^2}{\vec{k}^2} \leq 1.
\end{equation}
Consequently, we see that for a laminar plane wave of physical defects, the wave vector $k_\mu$ must be spacelike or lightlike and $v$ must be smaller than or equal to $\sqrt{k_\mu k^\mu}$.

\section{Energy--momentum}\label{sec:T}
In this section we will derive the energy--momentum tensor generated by an arbitrary configuration of physical defects in the limit that the distribution is continuous and all defect angles are small. This will tells us what conditions are imposed on the energy--momentum tensor by the requirement that the defects are physical (i.e. non-tachyonic). This puts limits on the kinds of models which can be found as the continuum limit of the piecewise linear model considered here.

In the linear weak field limit the energy--momentum tensor of a configuration of defects can be found by adding together the energy momentum tensors generated by the individual defects. Therefore, if $\hat{T}_{\mu\nu}\bb{\vec{p},\hat{d},\vec{v}}(x_\kappa)$ is the energy--momentum generated by a single defect with position $\vec{p}$, direction $\hat{d}$ and velocity $\vec{v}$ with unit energy density, then the total energy--momentum of a configuration of defects given by a density function $\rho(\vec{p},\hat{d},\vec{v})$ on $\cs$ is
\begin{equation}\label{eq:posT}
T[\rho]_{\mu\nu}(x_\kappa) = 
\int_\cs \!\!\!\!\md\vec{p}\,\md\hat{d}\,\md\vec{v}\;
\rho(\vec{p},\hat{d},\vec{v})\, \hat{T}_{\mu\nu}\bb{\vec{p},\hat{d},\vec{v}}(x_\kappa).
\end{equation}
Consequently, if we have an explicit expression for $\hat{T}_{\mu\nu}\bb{\vec{p},\hat{d},\vec{v}}(x_\kappa)$, we can compute the energy--momentum tensor for any configuration of defects. Alternatively, since the laminar plane waves form a complete basis for all configurations, the total energy--momentum of a configuration can also be obtained from the energy--momentum, $\hat{T}^{\lp}_{\mu\nu}\bb{k_\mu,\hat{d},v}(x_\kappa)$, of a laminar plane wave with wave vector $k_\mu$, direction $\hat{d}$ and perpendicular velocity $v$ through,
\begin{equation}
T[\bar\rho]_{\mu\nu}(x_\kappa) = 
\int_{\csp} \!\!\!\!\md k\,\md\hat{d}\,\md v\;
\bar\rho(k_\mu,\hat{d},v)\, \hat{T}^{\lp}_{\mu\nu}\bb{k_\mu,\hat{d},v}(x_\kappa).
\end{equation}

We will now derive the energy--momentum tensor $\hat{T}^{\lp}_{\mu\nu}\bb{k_\mu,\hat{d},v}(x_\kappa)$ generated by a single laminar plane wave. In \cite{hooft2008} 't Hooft derived the energy--momentum  tensor of a single stationary defect through the origin and directed along the $z$-axis, a result that was already well-known from the context of cosmic strings (see for example \cite{Brandenberger:2007ae}). With unit energy density the result is, 
\begin{equation}
\hat{T}_{\mu\nu}\bb{\vec{0},\hat{z},\vec{0}}(x_\kappa)=\begin{pmatrix}
1 & 0 & 0 & 0 \\
0 & 0 & 0 & 0 \\
0 & 0 & 0 & 0 \\
0 & 0 & 0 & -1 \\
\end{pmatrix}\delta(x)\delta(y).
\end{equation}
The dependence on the position $\vec{p}=(p_x,p_y,0)$\footnote{Remember that $\vec{p}$ should be perpendicular to $\hat{d}$.} can be obtained by performing appropriate shifts, which yields
\begin{equation}
\hat{T}_{\mu\nu}\bb{(p_x,p_y,0),\hat{z},\vec{0}}(x_\kappa)=\begin{pmatrix}
1 & 0 & 0 & 0 \\
0 & 0 & 0 & 0 \\
0 & 0 & 0 & 0 \\
0 & 0 & 0 & -1 \\
\end{pmatrix}\delta(x-p_x)\delta(y-p_y).
\end{equation}
By combining this result with equations \eqref{eq:lpw} and \eqref{eq:posT} we obtain the energy--momentum generated by a stationary laminar plane wave with wave vector $k_\mu=(0,\vec{k})$, direction $\hat{z}$ and zero velocity,
\begin{equation}\label{eq:statwave}
\begin{aligned}
\hat{T}^{\lp}_{\mu\nu}\bb{k_\lambda,\hat{z},0}(x_\kappa) &=
\int_{\cs}\!\!\!\! \md\vec{p}\,\md\hat{d}\,\md\vec{v}\; \rho^{\lp}[k_\lambda,\hat{z},0](\vec{p},\hat{d},\vec{v})\, \hat{T}_{\mu\nu}\bb{\vec{p},\hat{d},\vec{v}}(x_\kappa)\\
&=\int_{\vec{p}\perp\hat{z}}\!\!\!\! \md\vec{p}\;\hat{T}_{\mu\nu}\bb{\vec{p},\hat{z},\vec{0}}(x_\kappa)e^{2\pi\ii \vec{k}\cdot\vec{p}}\\
&=\begin{pmatrix}
1 & 0 & 0 & 0 \\
0 & 0 & 0 & 0 \\
0 & 0 & 0 & 0 \\
0 & 0 & 0 & -1 \\
\end{pmatrix}e^{2\pi\ii \vec{k}\cdot\vec{x}}.
\end{aligned}
\end{equation}

Any other laminar plane wave can be obtained by applying appropriate Lorentz transformation $\Lambda^\mu_\nu$. To write the result we first notice that the right hand side of equation to \eqref{eq:statwave} can be written as
\begin{equation}\label{eq:covar1}
\hh{u_\mu u_\nu + u^2 d_\mu d_\nu}e^{2\pi\ii k\cdot x},
\end{equation}
if we introduce the 4-vectors $u_\mu =  (1,0,0,0)$ and $d_\mu = (0,0,0,1)$.

After applying $\Lambda^\mu_\nu$ and a linear redefinition\footnote{The 4-vectors $u_\mu$ and $d_\mu$ span the plane of a single defect in the laminar plane wave. Applying a Lorentz transformation yields vectors spanning the plane of a defect in the transformed plane wave. However, these vectors will not correspond to the (non-covariant) parameters we introduced to describe the velocity and direction of the defect. To find a pair of 4-vectors $u'_\mu$  and $d'_\mu$ that span the same plane but correspond to the velocity $\vec{v}$ and direction $\hat{d}$ of the defect we need to do a linear transformation.} 
\begin{equation}
\begin{aligned}
d'_\nu &= \alpha  \Lambda^\mu_\nu d_\mu + \beta \Lambda^\mu_\nu  u_\nu,\\
u'_\mu &= \gamma \Lambda^\mu_\nu d_\mu + \delta \Lambda^\mu_\nu u_\mu
\end{aligned}
\end{equation}
such that 
\begin{equation}\label{cond:redef}
\begin{aligned}
d'_\mu d'^\mu&=1, & d'_0 &=0, \\
u'_\mu d'^\mu&=0, & u'_0 &=1,
\end{aligned}
\end{equation}
the direction $\hat{d}$ and velocity $\vec{v}$ of the new laminar plane wave can be found as
\begin{equation}
\begin{aligned}
u'_\mu &= (1,\vec{v}),\quad\text{and}\\
d'_\mu &= (0,\hat{d}).
\end{aligned}
\end{equation}
Applying the same Lorentz transformation and linear redefinition to equation \eqref{eq:covar1} yields the energy--momentum of the new laminar plane wave
\begin{equation}\label{eq:Tlp}
\hat{T}^{\lp}_{\mu\nu}[k_\lambda,\hat{d},v](x_\kappa) =
-\frac{1}{u'^2}\hh{u'_\mu u'_\nu + u'^2 d'_\mu d'_\nu}e^{2\pi\ii k\cdot x}.
\end{equation}
This gives us the energy--momentum for all laminar plane waves with $k_\mu k^\mu > 0$. Notice that a laminar plane wave of defects with wavevector $k_\mu$ only contributes to the Fourier mode of the energy--momentum tensor with wavevector $k_\mu$.  As a result the Fourier transform of the total energy--momentum of a configuration specified by the density function $\bar\rho$ on $\csp$ has the especially simple form,
\begin{equation}\label{eq:totalTk}
T_{\mu\nu}[\bar\rho](k_\lambda) = 
\int_{\hat{d}\perp\vec{k}} \!\!\!\!\md\hat{d}
	\!\!\int_0^{\sqrt{k_\lambda k^\lambda}}\s{-50}\md v \;
	 	\bar\rho(k_\lambda,\hat{d},v)\hat{T}^{pl}_{\mu\nu}[k_\lambda,\hat{d},v].
\end{equation}
The energy--momentum tensor for laminar plane wave with $k_\mu k^\mu =0$, can be found as a limit of equation \eqref{eq:Tlp}. There are two possibilities,
\begin{enumerate}
\item $k_\mu \rightarrow 0$ corresponding to a constant ``wave'' of defects with the same direction and velocity.
\item $u_\mu u^\mu \rightarrow 0$ corresponding to a laminar plane wave of lightlike defects.
\end{enumerate}
The first case is easy enough  to compute since equation \eqref{eq:Tlp} is regular in this limit. However, in the second limit the $1/u'^2$ factor blows up. This can fixed by noting that the normalization of $\hat{T}^{\lp}_{\mu\nu}$ is arbitrary, and we are therefore free to rescale it by a factor $-u'^2$ changing equation \eqref{eq:Tlp} to
\begin{equation}\label{eq:Tlp2}
\hat{T}^{\lp}_{\mu\nu}[k_\lambda,\hat{d},v](x_\kappa) =
\hh{u'_\mu u'_\nu + u'^2 d'_\mu d'_\nu}e^{2\pi\ii k\cdot x},
\end{equation}
which is regular in the limit that $u'^2$ goes to zero. This fixes $\hat{T}^{\lp}_{00}$ to be 1, which has the additional advantage of giving $\bar\rho$ the physical interpretation of the energy density present in a particular mode of laminar plane wave.

We now have all the ingredients we need to calculate the energy--momentum tensor of a general configuration of physical defects and can explore what conditions this will satisfy. In particular, we are interested in what conditions are imposed on the energy--momentum tensor by the restriction that all defects must be physical.

One immediate condition that we observe from equation \eqref{eq:Tlp} is that $k^\mu\hat{T}^{\lp}_{\mu\nu} = 0$ because $k^\mu d_\mu = k^\mu u_\mu = 0$. Consequently, equation \eqref{eq:totalTk} implies that the total energy--momentum must satisfy
\begin{equation}
k^\mu T_{\mu\nu}[\bar\rho] = 0,
\end{equation}
for any distribution $\bar\rho$ on $\csp$. That is, energy is conserved, as one expects from any reasonable physical theory.

Other conditions can be obtained by examining $T_{\mu\nu}[\bar\rho](k_\lambda)$ mode for mode. We have already observed that laminar plane waves with wave vector $k_\mu$ only contribute to modes of the energy--momentum tensor with the same wave vector. Therefore, since laminar plane waves of physical defects must have $k_\mu k^\mu \geq 0$, it is impossible for a configuration of defects to generate an energy--momentum tensor with Fourier modes with $k_\mu k^\mu < 0$.

For modes with $k_\mu k^\mu > 0$ we can assume by Lorentz invariance that, $k_\mu = (0,k,0,0)$. The direction $\hat{d}$, which must be perpendicular to $\vec{k}$, can then be parameterized by a single angle $\phi$ with $\phi=0$ corresponding to the direction of the $\hat{z}$-axis. The Fourier transform of equation \eqref{eq:Tlp2} then becomes,
\begin{equation}\label{eq:spacelikeT}
\hat{T}^{\lp}_{\mu\nu}[k_\kappa,\hat{d},v]=
\begin{pmatrix}
1 
	& 0 
		& \frac{v}{k} \cos\phi 
			 &  \frac{v}{k} \sin\phi \\
0 
	& 0 
		& 0
			& 0 \\
 \frac{v}{k} \cos\phi
	&  0
		&  \frac{v^2}{k^2} - \sin^2\phi
			& \cos\phi\sin\phi\\
  \frac{v}{k} \sin\phi
 	& 0
 		& \cos\phi\sin\phi
 			& \frac{v^2}{k^2} - \cos^2\phi\\
\end{pmatrix}.
\end{equation}
If we expand $\bar\rho(k_\mu,\hat{d},v)$ as
\begin{equation}\label{eq:expansion}
\bar\rho(k_\mu,\hat{d},v) =
\sum_{n=0}^\infty \frac{(2n+1)}{\pi}P_n (\tfrac{2v}{\sqrt{k^\mu k_\mu}}-1)
\bhh{r_{0n} + 2 \sum_{m=1}^\infty( r_{mn} \cos m\phi + \tilde{r}_{m n}\sin m\phi)},
\end{equation}
where the $P_n$ are Legendre polynomials and the coefficients $r_{mn}$ are understood to be functions of $k_\mu$, then the total contribution of the $k_\mu$-mode of $\bar\rho$ to the energy--momentum \eqref{eq:totalTk} becomes,
\begin{equation}\label{eq:totalTexp}
T_{\mu\nu}[\bar\rho](k_\mu) =
\begin{pmatrix}
2 r_{00} 
	& 0 
		& r_{11} +r_{10}
			 & \tilde{r}_{11} +\tilde{r}_{10} \\
0 
	& 0 
		& 0
			& 0 \\
r_{11}\!+\!r_{10}
	&  0
		& \tfrac{-1}{3}r_{00}\!+\!r_{01}\!+\!\tfrac{1}{3}r_{02}\!+\!r_{20}
			& \tilde{r}_{20}\\
\tilde{r}_{11}\!+\!\tilde{r}_{10}
 	& 0
 		&\tilde{r}_{20}
 			& \tfrac{-1}{3}r_{00}\!+\!r_{01}\!+\!\tfrac{1}{3}r_{02}\!-\!r_{20}\\
\end{pmatrix}.
\end{equation}
If we compare this to the most general form of a mode energy--momentum tensor with wave vector $k_\mu=(0,k,0,0)$ that satisfies $k^\mu T_{\mu\nu}=0$,
\begin{equation}
T_{\mu\nu}(k_\kappa) =
\begin{pmatrix}
 T_{00} 
	& 0 
		&T_{01}
			 & T_{02} \\
0 
	& 0 
		& 0
			& 0 \\
T_{01}
	&  0
		&T_{22}
			& T_{23}\\
T_{02}
 	& 0
 		&T_{23}
 			& T_{33}\\
\end{pmatrix},
\end{equation}
then we find that any such energy--momentum tensor can be generated by appropriate choices of the coefficients $r_{mn}$. We therefore find the physicality condition on the defects puts no further restrictions on the modes of the energy--momentum tensor with $k_\mu k^\mu > 0$.

In the special case that $k_\mu$ is lightlike, we can assume, without loss of generality, that $k_\mu= (\omega,\omega,0,0)$. In this limit  \eqref{eq:Tlp2} becomes,
\begin{equation}
\hat{T}^{\lp}_{\mu\nu}[k_\mu,\hat{d},v]=
\begin{pmatrix}
1 
	& 1 
		& 0 
			 & 0 \\
1
	& 1 
		& 0
			& 0 \\
0
	&  0
		& 0
			&0\\
 0
 	& 0
 		& 0
 			& 0\\
\end{pmatrix},
\end{equation}
and the total contribution of a distribution of defects $\bar\rho$ to the Fourier mode of the energy--momentum with $k_\mu= (\omega,\omega,0,0)$ becomes
\begin{equation}\label{eq:totalTexp2}
T_{\mu\nu}[\bar\rho](k_\mu)=
\begin{pmatrix}
r_{00} 
	& r_{00} 
		& 0 
			 & 0 \\
r_{00}
	& r_{00} 
		& 0
			& 0 \\
0
	&  0
		& 0
			&0\\
 0
 	& 0
 		& 0
 			& 0\\
\end{pmatrix}.
\end{equation}
Apparently additional restrictions apply to the lightlike modes of the energy--momentum tensor generated by a configuration of  physical defects. Not only do these modes have to be transverse, but they also cannot have any pressure or momentum perpendicular to their direction of propagation. This restriction can be formalized in the following way: For any lightlike mode of the energy--momentum tensor $T_{\mu\nu}(k_\kappa)$ and any lightlike vector $l^\mu$, the contraction $l^\mu T_{\mu\nu}(k_\kappa)$ is a non-spacelike vector. 

The other special case, $k_\mu=0$ is somewhat different since any direction $\hat{d}$ will be perpendicular to $\vec{k}=(0,0,0)$. The total contribution to the zero mode of the energy--momentum tensor will thus be found by integrating equation $\eqref{eq:Tlp2}$ over all mutually perpendicular $\vec{v}$ and $\hat{d}$. That is,
\begin{equation}
T_{\mu\nu}[\bar\rho](0) = -\int_{\vec{v}\perp\hat{d}} \md\vec{v}\md\hat{d}\; \bar\rho(0,\hat{d},\vec{v})(u_\mu u_\nu + u^2 d_\mu d_\nu).
\end{equation}
Since tensors of the form $u_\mu u_\nu + u^2 d_\mu d_\nu$ span the space of symmetric 2-tensors, we can conclude that any zero mode of the energy--momentum tensor may be produced by a configuration of defects. The condition that it contains only physical defects poses no further restrictions.

We therefore obtain the following restrictions that the energy-momentum tensor of a configuration of physical defects will satisfy in the continuum weak field limit
\begin{enumerate}
\item $k^\mu T_{\mu\nu}(k_\lambda) = 0$ for all $k_\mu$.
\item $T_{\mu\nu}(k_\lambda) = 0$ for all $k_\mu$ with $k^\mu k_\mu < 0$.
\item $l^\mu T_{\mu\nu}(k_\lambda)$ is a non-spacelike vector for all lightlike $k_\mu$ and $l_\mu$.
\end{enumerate}

The first and third condition hold for most physically reasonable theories. The first expresses conservation of energy--momentum, while the last is normally imposed as part of the null dominant energy condition which is employed in cosmology to ensure vacuum stability while allowing negative vacuum energy.\cite{CHT:2003}

The second condition is satisfied by various simple matter models used in general relativity, such as dusts. However, it is typically violated in classical wave like systems. For example, consider a standing wave solution of the Klein--Gordon equation, $\phi =  \cos(\omega t)\cos(\vec{k}\cdot\vec{x})$. Even if $\vec{k}^2 > \omega^2$, the energy--momentum tensor --- which behaves like the square of $\phi$ --- will have terms which are proportional to $\cos(2\omega t)$ and consequently will violate the second condition.

This indicates that the model cannot represent all types of matter at linear order. At this level all interactions are neglected, and we end up with a system that is very similar to a dust of non-interacting point particles. Beyond  linear order defect lines will collide in a non-trivial manner, as was discussed in \cite{meent:2010}. The energy--momentum tensor corresponding to the resolution of a collision will typically violate the second condition.

\section{Metric perturbations}\label{sec:h}
We now turn to the effect that a configuration of defects $\bar\rho(k_\mu,\hat{d},v)$ has on the metric. In the limit of weak fields the metric $g_{\mu\nu}$ can be separated in a static Minkowski background $\eta_{\mu\nu}$ and a small perturbation $h_{\mu\nu}$,
\begin{equation}
g_{\mu\nu}(x_\lambda) = \eta_{\mu\nu} + h_{\mu\nu}(x_\lambda).
\end{equation}
If we consider only the linear perturbations caused by the presence of a defect, then the total perturbation caused by a continuous distribution of defects can be found as the integral of the perturbations of individual components. Consequently, if $ \hat{h}^{\lp}_{\mu\nu}[k_\lambda,\hat{d},v](x_\kappa)$ is the perturbation of the metric caused by a laminar plane wave  with wave vector $k_\mu$, direction $\hat{d}$, and perpendicular velocity $v$, then the total perturbation caused by a configuration $\bar\rho$ is given by,
\begin{equation}\label{eq:totalH}
h_{\mu\nu}[\bar\rho](x_\kappa) = 
\int_{\csp} \!\!\!\!\md k\,\md\hat{d}\,\md v\;
\bar\rho(k_\lambda,\hat{d},v)\, 
\hat{h}^{\lp}_{\mu\nu}\bb{k_\lambda,\hat{d},v}(x_\kappa).
\end{equation}
Therefore, if we know $\hat{h}^{\lp}_{\mu\nu}\bb{k_\lambda,\hat{d},v}$ for any combination of the parameters $(k_\lambda,\hat{d},v)$, than $h_{\mu\nu}[\bar\rho]$ can be computed for any configuration $\bar\rho$. 

To calculate $\hat{h}^{\lp}_{\mu\nu}\bb{k_\lambda,\hat{d},v}(x_\kappa)$ we fix its gauge freedom with the condition $\partial^\mu h_{\mu\nu} =0$.\footnote{There is some residual gauge freedom for the lightlike modes of the metric perturbation as we will discuss later on.} With this choice the linearized Einstein equation becomes (in its Fourier transformed form),
\begin{equation}\label{eq:lineinstein}
T_{\mu\nu}(k_\kappa)= 2\pi^2k^2\bhh{h_{\mu\nu}-h(\eta_{\mu\nu}-\frac{k_\mu k_\nu}{k^2})},
\end{equation}
where $h$ is the trace of $h_{\mu\nu}$.

When $k^\mu k_\mu \neq 0$, equation \eqref{eq:lineinstein} can be inverted to obtain the linear metric perturbation as a function of the energy--momentum tensor.  In particular, if  $\hat{T}^{\lp}_{\mu\nu}\bb{k_\lambda,\hat{d},v}$ is the Fourier mode of the energy--momentum tensor generated by a laminar plane wave, then the metric perturbation generated by that laminar plane wave is given by a single Fourier mode,
\begin{equation}
 \hat{h}^{\lp}_{\mu\nu}[k_\lambda,\hat{d},v]= \frac{1}{2\pi^2 k^2}
 \bhh{\delta_\mu^\alpha\delta_\nu^\beta -\hf (\eta_{\mu\nu}-\tfrac{k_\mu k_\nu}{k^2})\eta^{\alpha\beta}}\hat{T}^{\lp}_{\alpha\beta}[k_\lambda,\hat{d},v].
\end{equation}
We can therefore study the effects of a distribution of defects $\bar\rho(k_\mu,\phi,v)$ on a per mode basis. By plugging in the $\hat{T}^{\lp}_{\mu\nu}[k_\lambda,\hat{d},v]$ from equation \eqref{eq:Tlp2}, we find that, 
\begin{equation}\label{eq:metricT}
\hat{h}^{\lp}_{\mu\nu}[k_\lambda,\hat{d},v]= \frac{1}{2\pi^2 k^2}\bhh{u_\mu u_\nu - u^2 (\eta_{\mu\nu}-d_\mu d_\nu -\frac{k_\mu k_\nu}{k^2})},
\end{equation}
where $u_\mu = (1,\vec{v})$ and $d_\mu = (0,\hat{d})$.

When  $k^\mu k_\mu > 0$,  we can assume by Lorentz invariance that  $k_\mu = (0,k,0,0)$.  Parameterizing $\hat{d}$ as $(0,-\sin\phi,\cos\phi)$ and $\vec{v}$ as $(0,v\cos\phi,v\sin\phi)$, we obtain the explicit expression,
\begin{equation}\label{eq:spacelikeh}
\hat{h}^{\lp}_{\mu\nu}[k_\lambda,\hat{d},v]=\frac{1}{2\pi^2 k^2}\begin{pmatrix}
\frac{v^2}{k^2}
	& 0 
		& \frac{v}{k} \cos\phi 
			 & \frac{v}{k} \sin\phi \\
0 
	& 0 
		& 0
			& 0 \\
\frac{v}{k} \cos\phi
	&  0
		&  \cos^2\phi
			& \cos\phi\sin\phi\\
\frac{v}{k} \sin\phi
 	& 0
 		& \cos\phi\sin\phi
 			& \sin^2\phi\\
\end{pmatrix}.
\end{equation}

The total metric perturbation caused by a configuration of defects given by a distribution $\bar\rho$ on $\csp$ can be obtained from $\hat{h}^{\lp}_{\mu\nu}[k_\lambda,\hat{d},v]$ through the integral,
\begin{equation}\label{eq:totalHk}
h_{\mu\nu}[\bar\rho](k_\lambda) = 
\int\md\hat{d}\,\md v\;
\bar\rho(k_\lambda,\hat{d},v)\, 
\hat{h}^{\lp}_{\mu\nu}\bb{k_\lambda,\hat{d},v}.
\end{equation}
Performing this integral in the case that $k_\mu=(0,k,0,0)$ and applying the expansion of $\bar\rho$ as given in \eqref{eq:expansion} yields,
\begin{equation}\label{eq:spacelikeHdist}
h_{\mu\nu}[\bar\rho](k_\kappa)=
\frac{1}{2 \pi^2 k^2}\begin{pmatrix}
\tfrac{2}{3}r_{00}+r_{01}+\tfrac{1}{3}r_{02}
	&\s{10} 0\s{10} 
		&\s{10}  r_{10} + r_{11}\s{10} 
			 & \tilde{r}_{10} + \tilde{r}_{11}\\
0 
	& 0 
		& 0
			& 0 \\
r_{10} + r_{11}
	&  0
		& r_{00} + r_{20}
			& \tilde{r}_{20}\\
 \tilde{r}_{10} + \tilde{r}_{11}
 	& 0
 		& \tilde{r}_{20}
 			& r_{00} - r_{20}\\
\end{pmatrix}.
\end{equation}

When $k^\mu k_\mu = 0$ the linearized Einstein equation \eqref{eq:lineinstein} cannot be inverted and equation \eqref{eq:metricT} cannot be applied directly.  However, we may obtain the metric perturbation for these modes as a limiting case of the modes with $k^\mu k_\mu > 0$. 

In the case that $k_\mu$ becomes lightlike we can assume due to Lorentz invariance that it goes to $k_\mu = (\omega,\omega,0,0)$. Such a laminar plane wave is the limit of waves with momentum $k_\mu = (\omega,\kappa,0,0)$ as $\kappa \rightarrow \omega$. Since physicality requires that $0\leq v \leq \sqrt{\kappa^2 -\omega^2}$, $v$ must simultaneously  go to zero. Simply setting $v=0$ and using $k_\mu = (\omega,\kappa,0,0)$, $u_\mu = (1,\omega/\kappa,0,0)$, and $d_\mu = (0,0,-\sin\phi,\cos\phi)$ in equation \eqref{eq:metricT} yields,
\begin{equation}\label{eq:v0h}
\hat{h}^{\lp}_{\mu\nu}[k_\lambda,\phi,0]=
-\frac{1}{2\pi^2 \kappa^2}\begin{pmatrix}
0
	& 0 
		& 0
			 & 0 \\
0 
	& 0 
		& 0
			& 0 \\
0
	&  0
		&  \cos^2\phi
			& \cos\phi\sin\phi\\
 0
 	& 0
 		& \cos\phi\sin\phi
 			& \sin^2\phi\\
\end{pmatrix}.
\end{equation}
Since only the pre-factor depends on $\kappa$ the limit as $\kappa$ goes to $\omega$  is straight forward.

Letting $v$ go to zero by another route will lead to a result that (in the $\kappa \rightarrow \omega$ limit) differs from the above by a term of the following form
\begin{equation}\label{eq:vah}
\begin{pmatrix}
\xi_1
	& \xi_1 
		& \xi_2
			 & \xi_3 \\
\xi_1
	& \xi_1 
		& \xi_2
			& \xi_3 \\
\xi_2
	&  \xi_2
		&  0
			& 0\\
 \xi_3
 	& \xi_3
 		& 0
 			&0\\
\end{pmatrix},
\end{equation}
where the $\xi_i$ are arbitrary (possibly infinite) parameters. 

Such a contribution can be gauged away.  Under an infinitesimal coordinate transformation given by a vector field $\xi_\mu$, the  $h_{\mu\nu}$ transforms as,
\begin{equation}
 h_{\mu\nu} \to h_{\mu\nu} + \partial_\mu\xi_\nu +\partial_\nu\xi_\mu.
\end{equation}
The gauge condition $\partial^\mu h_{\mu\nu} =0$ implies that
\begin{equation}
k^\mu k_\mu \xi_{\nu}(k_\lambda)+ k_\nu k^\mu \xi_{\mu}(k_\lambda)=0.
\end{equation}
This completely fixes $\xi_\mu(k_\lambda)$ for $k^\mu k_\mu \neq 0$. However, when $k^\mu k_\mu= 0$ it only implies that $k^\mu \xi_{\mu}(k_\lambda) = 0$. The residual gauge transformation subject to that condition for $k_\mu = (\omega,\omega,0,0)$ takes the form of equation \eqref{eq:vah}. Contributions of that form can therefore be gauged away, and we are free to adopt \eqref{eq:v0h} as the gauge fixed form of $\hat{h}^{\lp}_{\mu\nu}[(\omega,\omega,0,0),\phi,0]$.

Inserting \eqref{eq:v0h} in the integral \eqref{eq:totalHk} and using the expansion \eqref{eq:expansion} for $\bar\rho$ yields the lightlike modes of the metric perturbation caused by a configuration of defects given by a distribution $\bar\rho$
\begin{equation}\label{eq:lightlikeHdist}
h_{\mu\nu}[\bar\rho](\omega,\omega,0,0)=
\frac{1}{2 \pi^2 \omega^2}\begin{pmatrix}
0
	& 0 
		&0
			 & 0\\
0 
	& 0 
		& 0
			& 0 \\
0
	&  0
		& r_{00} + r_{20}
			& \tilde{r}_{20}\\
0
 	& 0
 		& \tilde{r}_{20}
 			& r_{00} - r_{20}\\
\end{pmatrix}.
\end{equation}

In the case that $k_\mu \rightarrow 0$, the equation \eqref{eq:metricT} diverges. This signals a breakdown of the linear perturbation approach in the limit of constant fields. This is not unexpected, since constant non-zero energy densities will typically lead to non-trivial effects on the global level with respect to the topology or causal structure. 

\section{Gravitational waves}\label{sec:gravwave}
In the previous two sections we have obtained the energy--momentum and metric perturbation caused by a continuous distribution of defects at the linear level. At the microscopic level of individual defects the energy--momentum tensor completely fixed the  metric (up to gauge transformations) due to the ad hoc rule we imposed that the vacuum should be completely flat. We are now able to answer the question whether this property persists in the continuum limit.

Since we are considering the linear limit, it is enough to consider what additional metric structure maybe present for a configuration with zero energy--momentum. In section \ref{sec:T} we obtained a complete expression for the energy--momentum of a configuration of defects described by a distribution $\bar\rho$ on $\csp$. Requiring that this expression (equations \eqref{eq:totalTexp} and \eqref{eq:totalTexp2}) vanishes implies the following conditions on the coefficients $r_{nm}(k_\mu)$ of $\bar\rho$ in the expansion \eqref{eq:expansion};
\begin{align}
\left. \begin{aligned}
 r_{00}(k_\mu) &= 0,	& r_{01}(k_\mu)+ \tfrac{1}{3}r_{02}(k_\mu)&= 0,\\
 \tilde{r}_{20}(k_\mu) &= 0,	& r_{11}(k_\mu)+ r_{10}(k_\mu)&= 0,\\
 r_{20}(k_\mu) &= 0,	& \tilde{r}_{11}(k_\mu)+ \tilde{r}_{10}(k_\mu)&= 0,
\end{aligned}\right\}&\quad\text{for $k^\mu k_\mu >0$, and}\\
\left.\begin{aligned}
r_{00}(k_\mu) & = 0,
\end{aligned}\right\}&\quad\text{for $k^\mu k_\mu =0$.}
\end{align}

The metric perturbation caused by a vacuum configuration of defects can now be found by applying these conditions to the complete expressions \eqref{eq:spacelikeHdist} and \eqref{eq:lightlikeHdist}, found in section \ref{sec:h} for the metric perturbation caused by a configuration of defects. This yields that for $k^\mu k_\mu >0$, $h_{\mu\nu}(k_\lambda)$ vanishes when $T_{\mu\nu}$ vanishes, as one would expect since the linearized Einstein equation is invertible for $k^\mu k_\mu \neq 0$.

However, for $k^\mu k_\mu =0$, the requirement that $T_{\mu\nu}(k_\lambda)$ vanishes only fixes $r_{00}(k_\mu)$  to be zero. The coefficients $r_{20}(k_\mu)$ and $\tilde{r}_{20}(k_\mu)$ are  unconstrained. Consequently, for each lightlike $k_\mu$ there exists a two parameter family of vacuum metric structures. When  $k_\mu= (\omega,\omega,0,0)$ these are described by 
\begin{equation}\label{eq:gravwaves}
h_{\mu\nu}[\bar\rho](\omega,\omega,0,0)=
\frac{1}{2 \pi^2 \omega^2}\begin{pmatrix}
0
	& 0 
		&0
			 & 0\\
0 
	& 0 
		& 0
			& 0 \\
0
	&  0
		& r_{20}
			& \tilde{r}_{20}\\
0
 	& 0
 		& \tilde{r}_{20}
 			&  - r_{20}\\
\end{pmatrix},
\end{equation}
where we immediately recognize $r_{20}$ and $\tilde{r}_{20}$ as the coefficients of the familiar $+$ and $\times$ polarizations of gravitational waves. 

This answers the question we posed in the introduction of whether any of the vacuum structure of general relativity would be recovered in the continuum limit of our model. The answer turns out to be affirmative. In fact, we find that all vacuum (i.e. Ricci flat) metrics that are available in linearized general relativity may be found as the continuum limit of a sequence of configurations of physical defects.

\section{Conclusions and Outlook}
We have studied continuous distributions of physical defects. In the limit of weak fields, considering only linear contributions, the energy--momentum tensor of such a distribution turns out to satisfy certain conditions. These conditions can be expressed as follows for the Fourier transform of the energy--momentum tensor,
 \begin{enumerate}
\item $k^\mu T_{\mu\nu}(k_\lambda) = 0$ for all $k_\mu$.
\item $T_{\mu\nu}(k_\lambda) = 0$ for all $k_\mu$ with $k^\mu k_\mu < 0$.
\item $l^\mu T_{\mu\nu}(k_\lambda)$ is a non-spacelike vector for all lightlike $k_\mu$ and $l_\mu$.
\end{enumerate}
The first and third conditions are satisfied by most reasonable classical theories. The second condition, shows an inability to model wave-like phenomenon at the linear order. This property is shared by other non-interacting matter models, like dusts. Beyond linear order line defects collide in a non-trivial manner causing violations of this second condition. It would be very interesting to see if any restrictions remain once interactions are included.

The metric perturbation caused by a configuration of defects turns out to be mostly fixed by its energy--momentum. That is if two continuous distributions of defects have the same energy--momentum tensor, they also produce the same linear metric perturbation. At least for most modes. 

The lightlike modes of the metric perturbation form an exception, as they are only partially fixed by the energy--momentum tensor. The metric perturbations of two configurations with the same energy--momentum may differ by a transverse traceless lightlike mode, i.e. by a gravitational plane wave. We thus see that the piecewise flat model of propagating defects --- even though it does not contain any a priori vacuum structure --- recovers the vacuum structure of general relativity in the continuum limit, at least at the linearized level.

The analysis in this article is possible because the description of the model of propagating line defects simplifies dramatically in the linear limit. Of course, it would be interesting to see what happens to these results if we go beyond the linear level. Besides seeing what restrictions may persist on the energy--momentum tensor obtained in the continuum limit, it would be particularly interesting to see whether the non-linear vacuum structures of general relativity --- like black hole horizons --- are recovered.

Going beyond the linear level is difficult however. Not only do the descriptions of the defect lines become intertwined with the presence of other defect lines, but we also need to account for collisions of the line defects. The resolution of such collisions was discussed in our previous paper \cite{meent:2010}, where we saw that these resolutions are not unique (and therefore require extra physical input) and sometimes require the appearance of superluminal nodes, which go against the spirit of the model which insists that all features propagate at subluminal speeds.

A general analysis of configurations, as done in this paper, therefore seems unlikely at the non-linear level. It might however be possible to produce specific examples of non-linear vacuum structure by taking intuition from the linear limit. For example, the linear analysis showed that only lightlike defects moving in a single direction are needed to form a gravitational plane wave. This suggests that the constituent defects of a gravitational wave do not collide with each other.  This simplifies the description of such a solution. One might therefore hope to form a family of exact piecewise flat solutions, which have zero energy--momentum when average over some region and which approximate a family of exact gravitational plane wave solutions like the one found by Bondi et al.\cite{bondi:1959} Some progress has been made in this direction.

Similarly, one might hope to use spherical symmetry to produce a piecewise flat vacuum approximation to a Schwarzschild black hole.

\section*{Acknowledgments}
The author would like to thank his advisor, Gerard 't Hooft, for many valuable discussions.
\bibliographystyle{utcaps}
\bibliography{gravwave}
\end{document}